\documentclass{elsart}
\usepackage{ifpdf}
\usepackage{graphicx,amssymb,lineno,epsfig,subfig}
\usepackage{subfig}
\usepackage{indentfirst}

\ifpdf
\usepackage[%
  pdftitle={Instructions for use of the document class
    elsart},%
  pdfauthor={Simon Pepping},%
  pdfsubject={The preprint document class elsart},%
  pdfkeywords={instructions for use, elsart, document class},%
  pdfstartview=FitH,%
  bookmarks=true,%
  bookmarksopen=true,%
  breaklinks=true,%
  colorlinks=true,%
  linkcolor=blue,anchorcolor=blue,%
  citecolor=blue,filecolor=blue,%
  menucolor=blue,pagecolor=blue,%
  urlcolor=blue]{hyperref}
\else
\usepackage[%
  breaklinks=true,%
  colorlinks=true,%
  linkcolor=blue,anchorcolor=blue,%
  citecolor=blue,filecolor=blue,%
  menucolor=blue,pagecolor=blue,%
  urlcolor=blue]{hyperref}
\fi

\renewcommand{\theequation}{\arabic{section}.\arabic{equation}}

\def\@subjclass{}
\makeatletter
\def\elsartstyle{%
    \def\normalsize{\@setfontsize\normalsize\@xiipt{14.5}}
    \def\small{\@setfontsize\small\@xipt{13.6}}
    \let\footnotesize=\small
    \def\large{\@setfontsize\large\@xivpt{18}}
    \def\Large{\@setfontsize\Large\@xviipt{22}}
    \skip\@mpfootins = 18\p@ \@plus 2\p@
    \normalsize
} \@ifundefined{square}{}{\let\Box\square} \makeatother
\usepackage{setspace}
\usepackage{indentfirst}

\begin{document}
\begin{frontmatter}
\title{Soliton and periodic wave solutions to the osmosis K(2, 2) equation}
\author{Jiangbo Zhou\corauthref{cor}},
\corauth[cor]{Corresponding author. } \ead{zhoujiangbo@yahoo.cn}
\author{Lixin Tian},
\author{Xinghua Fan}
\address{Nonlinear Scientific Research Center, Faculty of Science, Jiangsu
University, Zhenjiang, Jiangsu 212013, China}
\begin{abstract} In this paper, two types of traveling wave solutions to the
osmosis K(2, 2) equation
 \[u_t +(u^2)_x -(u^2)_{xxx}=0 \]
 are investigated. They are characterized by
two parameters. The expresssions for the soliton and periodic wave
solutions are obtained.
\end{abstract}


\end{frontmatter}
\section{Introduction}
 \setcounter {equation}{0}

In 1993, Rosenau and Hyman \cite {1} introduced a genuinely
nonlinear dispersive equation, a special type of KdV equation, of
the form
\begin{equation}
\label {eq1.1}u_t +a(u^n)_x + (u^n)_{xxx}=0, n>1,
\end{equation}
where $a$ is a constant and both the convection term $(u^n)_x$ and
the dispersion effect term $(u^n)_{xxx}$ are nonlinear. These
equations arise in the process of understanding the role of
nonlinear dispersion in the formation of structures like liquid
drops. Rosenau and Hyman derived solutions called compactons to
Eq.(\ref{eq1.1}) and showed that while compactons are the essence of
the focusing branch where $a>0$, spikes, peaks, and cusps are the
hallmark of the defocusing branch where $a<0$  which also supports
the motion of kinks. Further, the negative branch, where $a<0$, was
found to give rise to solitary patterns having cusps or infinite
slopes. The focusing branch and the defocusing branch represent two
different models, each leading to a different physical structure.
Many powerful methods were applied to construct the exact solutions
to Eq.(\ref{eq1.1}), such as Adomain method \cite {2}, homotopy
perturbation method \cite {3}, Exp-function method \cite {4},
variational iteration method \cite {5}, variational method \cite {6,
7}. In \cite {8}, Wazwaz studied a generalized forms of the
Eq.(\ref{eq1.1}), that is $mK(n,n)$ equations and defined by
\begin{equation}
 \label {eq1.2} u^{n-1}u_t +a(u^n)_x+ b(u^n)_{xxx}=0, n>1,
\end{equation}
where $a, b$ are constants. He showed how to construct compact and
noncompact solutions to Eq.(\ref{eq1.2}) and discussed it in higher
dimensional spaces in  \cite {9}. Chen et al.  \cite {10} showed how
to construct the general solutions and some special exact solutions
to Eq.(\ref{eq1.2}) in higher dimensional spatial domains. He et al.
\cite {11} considered the bifurcation behavior of travelling wave
solutions to Eq.(\ref{eq1.2}). Under different parametric
conditions, smooth and non-smooth periodic wave solutions, solitary
wave solutions and kink and anti-kink wave solutions were obtained.
Yan \cite {12} further extended Eq.(\ref{eq1.2}) to be a more
general form
\begin{equation}
 \label {eq1.3} u^{m-1}u_t +a(u^n)_x+ b(u^k)_{xxx}=0, nk\neq 1,
\end{equation}
And using some direct ansatze, some abundant new compacton
solutions, solitary wave solutions and periodic wave solutions to
Eq.(\ref{eq1.3}) were obtained. By using some transformations, Yan
\cite {13} obtained some Jacobi elliptic function solutions to
Eq.(\ref{eq1.3}). Biswas \cite {14} obtained 1-soliton solution of
equation with the generalized evolution term
 \begin{equation}
 \label {eq1.4} (u^l)_t +a(u^m)u_x+ b(u^n)_{xxx}=0,
\end{equation}
where $a, b$ are constants, while $l, m$ and $n$ are positive
integers. Zhu et al. \cite {15} applied the decomposition method and
symbolic computation system to develop some new exact solitary wave
solutions to the $K(2, 2, 1)$ equation
\begin{equation}
 \label {eq1.5}  u_t +(u^2)_x - (u^2)_{xxx}+u_{xxxxx}=0,
\end{equation}
 and the $K(3, 3, 1)$ equation
\begin{equation}
 \label {eq1.6}  u_t +(u^3)_x - (u^3)_{xxx}+u_{xxxxx}=0.
\end{equation}
Recently, Xu and Tian \cite{16} introduced the osmosis $K(2,2)$
equation
\begin{equation}
 \label {eq1.7}  u_t +(u^2)_x - (u^2)_{xxx}=0,
\end{equation}
were the positive convection term $(u^2)_x$ means the convection
moves along the motion direction, and the negative dispersive term
$(u^2)_{xxx}$ denotes the contracting dispersion. They obtained the
peaked solitary wave solution and the periodic cusp wave solution to
Eq.(\ref{eq1.7}). In \cite{17}, the authors obtained the smooth
soliton solutions to Eq.(\ref{eq1.7}). In this paper, following
Vakhnenko and Parkes's strategy \cite{18, 19} we continue to
investigate the traveling wave solutions to Eq.(\ref{eq1.7}) and
obtain soliton and periodic wave solutions. Our work in this paper
covers and extends the results in \cite{16, 17} and may help people
to know deeply the described physical process and possible
applications of the osmosis K(2, 2) equation.

The remainder of this paper is organized as follows. In Section 2,
for completeness and readability, we repeat Appendix A in \cite{19},
which discuss the solutions to a first-order ordinary differential
equaion. In Section 3, we show that, for travelng wave solutions,
Eq.(\ref{eq1.7}) may be reduced to a first-order ordinary
differential equation involving two arbitrary integration constants
$a$ and $b$. We show that there are four distinct periodic solutions
corresponding to four different ranges of values of $a$ and
restricted ranges of values of $b$. A short conclusion is given in
Section 4.

\section{Solutions to a first-order ordinary differential
equaion}
 \setcounter {equation}{0}

This section is due to Vakhnenko and Parkes (see Appendix A in
\cite{19}). For completeness and readability, we state it in the
following.

Consider solutions to the following ordinary
 differential equation
\begin{equation}
\label{eq2.1}
 (\varphi\varphi_\xi)^2=\varepsilon^2 f(\varphi),
\end{equation}
where
\begin{equation}
\label{eq2.2}
f(\varphi)=(\varphi-\varphi_1)(\varphi-\varphi_2)(\varphi_3-\varphi)(\varphi_4-\varphi),
\end{equation}
and $\varphi_1$, $\varphi_2$, $\varphi_3$, $\varphi_4$ are chosen to
be real constants with $\varphi_1\leq \varphi_2\leq \varphi\leq
\varphi_3 \leq \varphi_4$.

Following \cite{20} we introduce $\zeta$ defined by
\begin{equation}
\label{eq2.3} \frac{d\xi}{d\zeta}=\frac{\varphi}{\varepsilon},
\end{equation}
so that Eq.(\ref{eq2.1}) becomes
\begin{equation}
\label{eq2.4} (\varphi_\zeta)^2=f(\varphi).
\end{equation}

Eq.(\ref{eq2.4}) has two possible forms of solution. The first form
is found using result 254.00 in \cite{21}. Its parametric form is
\begin{equation}
\label{eq2.5} \left\{ {\begin{array}{l}
\varphi = \frac{\textstyle \varphi_2-\varphi_1 n \mathrm{sn}^2(w|m) }{\textstyle 1-n \mathrm{sn}^2(w|m)} ,\\
 \xi = \frac{\displaystyle 1}{\displaystyle \varepsilon
 p}(w\varphi_1+(\varphi_2-\varphi_1)\Pi(n;w|m)),
\\
 \end{array}} \right.
\end{equation}
with $w$ as the parameter, where
\begin{equation}
\label{eq2.6}
m=\frac{(\varphi_3-\varphi_2)(\varphi_4-\varphi_1)}{(\varphi_4-\varphi_2)(\varphi_3-\varphi_1)},
p=\frac{1}{2}\sqrt{(\varphi_4-\varphi_2)(\varphi_3-\varphi_1)},
w=p\zeta,
\end{equation}
and
\begin{equation}
\label{eq2.7} n=\frac{\varphi_3-\varphi_2}{\varphi_3-\varphi_1}.
\end{equation}
In (\ref{eq2.5}) $\mathrm{sn}(w|m)$ is a Jacobian elliptic function,
where the notation is as used in Chapter 16 of \cite{22}.
$\Pi(n;w|m)$ is the elliptic integral of the third kind and the
notation is as used in Section 17.2.15 of \cite{22}.

The solution to Eq.(\ref{eq2.1}) is given in parametric form by
(\ref{eq2.5}) with $w$ as the parameter. With respect to $w$,
$\varphi$ in (\ref{eq2.5}) is periodic with period $2K(m)$, where
$K(m)$ is the complete elliptic integral of the first kind. It
follows from (\ref{eq2.5}) that the wavelength $\lambda$ of the
solution to (\ref{eq2.1}) is
\begin{equation}
\label{eq2.8} \lambda = \frac{\displaystyle 2}{\displaystyle
 \varepsilon p}|\varphi_1K(m)+(\varphi_2-\varphi_1)\Pi(n|m)|.
 \end{equation}
where $\Pi(n|m)$ is the complete elliptic integral of the third
kind.

When $\varphi_3=\varphi_4$, $m=1$, (\ref{eq2.5}) becomes
\begin{equation}
\label{eq2.9} \left\{ {\begin{array}{l}
\varphi = \frac{\textstyle \varphi_2-\varphi_1 n \tanh^2w }{\textstyle 1-n  \tanh^2w} ,\\
 \xi = \frac{\textstyle 1}{\textstyle \varepsilon
 }(\frac{\textstyle w\varphi_3}{\textstyle p}-2\tanh^{-1}(\sqrt{n}\tanh
 w)).
\\
 \end{array}} \right.
\end{equation}

The second form of the solution to Eq.(\ref{eq2.4}) is found using
result 255.00 in \cite{21}. Its parametric form is

\begin{equation}
\label{eq2.10}  \left\{ {\begin{array}{l}
\varphi = \frac{\textstyle \varphi_3-\varphi_4 n \mathrm{sn}^2(w|m) }{\textstyle 1-n \mathrm{sn}^2(w|m)} ,\\
 \xi = \frac{\displaystyle 1}{\displaystyle \varepsilon
 p}(w\varphi_4-(\varphi_4-\varphi_3)\Pi(n;w|m)),
\\
 \end{array}} \right.
\end{equation}
where $m, p, w$ are as in (\ref{eq2.6}), and
\begin{equation}
\label{eq2.11} n=\frac{\varphi_3-\varphi_2}{\varphi_4-\varphi_2}.
\end{equation}

The solution to Eq.(\ref{eq2.1}) is given in parametric form by
(\ref{eq2.10}) with $w$ as the parameter. The wavelength $\lambda$
of the solution to (\ref{eq2.1}) is
\begin{equation}
\label{eq2.12} \lambda = \frac{\displaystyle 2}{\displaystyle
 \varepsilon p}|\varphi_4K(m)-(\varphi_4-\varphi_3)\Pi(n|m)|.
 \end{equation}

When $\varphi_1=\varphi_2$, $m=1$, (\ref{eq2.10}) becomes
\begin{equation}
\label{eq2.13} \left\{ {\begin{array}{l}
\varphi = \frac{\textstyle \varphi_3-\varphi_4 n \tanh^2w }{\textstyle 1-n  \tanh^2w} ,\\
 \xi = \frac{\textstyle 1}{\textstyle \varepsilon
 }(\frac{\textstyle w\varphi_2}{\textstyle p}+2\tanh^{-1}(\sqrt{n}\tanh
 w)).
\\
 \end{array}} \right.
\end{equation}

\section{Solitary and periodic wave solutions to Eq.(\ref{eq1.7})}
 \setcounter {equation}{0}

Eq.(\ref{eq1.7}) can also be written in the form
\begin{equation}
\label{eq3.1}  u_t +2uu_x - 6u_xu_{xx}-2uu_{xxx}=0.
 \end{equation}
Let $u=\varphi(\xi )+c$ with $\xi = x - ct$ be a traveling wave
solution to Eq.(\ref{eq3.1}), then it follows that
\begin{equation}
\label{eq3.2}  - c\varphi_\xi + 2\varphi \varphi_\xi-6\varphi
_\xi\varphi_{\xi\xi} -2\varphi \varphi _{\xi\xi\xi}=0,
 \end{equation}
where $\varphi_\xi$ is the derivative of function $\varphi$ with
respect to $\xi$.

Integrating (\ref{eq3.2}) twice with respect to $\xi$ yields
 \begin{equation}
\label{eq3.3}
(\varphi\varphi_\xi)^2=\frac{1}{4}(\varphi^4-\frac{4c}{3}\varphi^3+a
\varphi^2+b),
 \end{equation}
where $a$ and $b$ are two arbitrary integration constants.

Eq.(\ref{eq3.3}) is in the form of Eq.(\ref{eq2.1}) with
$\varepsilon=\frac{\textstyle1}{\textstyle2}$ and
$f(\varphi)=(\varphi^4-\frac{\textstyle4c}{\textstyle3}\varphi^3+a
\varphi^2+b)$. For convenience we define $g(\varphi)$ and
$h(\varphi)$ by
 \begin{equation}
\label{eq3.4} f(\varphi)=\varphi^2g(\varphi)+b, \  \mbox{where} \
g(\varphi)=\varphi^2-\frac{4c}{3}\varphi+a,
 \end{equation}
 \begin{equation}
\label{eq3.5} f'(\varphi)=2\varphi h(\varphi), \ \mbox{where} \
h(\varphi)=2\varphi^2-2c \varphi+a,
 \end{equation}
and define $\varphi_L$, $\varphi_R$, $b_L$, and $b_R$ by
 \begin{equation}
\label{eq3.6} \varphi_L=\frac{\textstyle 1}{\textstyle
2}(c-\sqrt{c^2-2a}), \varphi_R=\frac{\textstyle 1}{\textstyle
2}(c+\sqrt{c^2-2a}),
 \end{equation}
 \begin{equation}
\label{eq3.7} b_L=-\varphi_L^2g(\varphi_L)=\frac{\textstyle
a^2}{\textstyle 4}-\frac{\textstyle 1 }{\textstyle
2}c^2a+\frac{\textstyle c^4}{\textstyle 6}-\frac{\textstyle
1}{\textstyle 6}(c^3-2ac)\sqrt{c^2-2a},
 \end{equation}
 \begin{equation}
 \label{eq3.8} b_R=-\varphi_L^2g(\varphi_L)=\frac{\textstyle
a^2}{\textstyle 4}-\frac{\textstyle 1 }{\textstyle
2}c^2a+\frac{\textstyle c^4}{\textstyle 6}+\frac{\textstyle
1}{\textstyle 6}(c^3-2ac)\sqrt{c^2-2a}.
 \end{equation}
 Obviously, $\varphi_L$, $\varphi_R$ are the roots of $h(\varphi)=0$.

Without loss of generality, we suppose the wave speed $c>0$. In the
following, suppose that $a<\frac{\textstyle c^2 }{\textstyle 2}$ and
$a\neq 0$ for each value $c>0$, such that
 $f(\varphi)$ has three distinct stationary points: $\varphi_L$,
 $\varphi_R$, $0$ and comprise two minimums separated by a maximum.
 Under this assumption, Eq.(\ref{eq1.7}) has periodic and solitary wave
solutions that have different analytical forms depending on the
values of $a$ and $b$ as follows:

(1) $a<0$

In this case $\varphi_L<0<\varphi_R$ and
$f(\varphi_L)>f(\varphi_R)$. For each value $a<0$ and $0<b<b_L$ (a
corresponding curve of $f(\varphi)$ is shown in Fig.1(a)), there are
periodic loop-like solutions to Eq.(\ref{eq3.3}) given by
(\ref{eq2.10}) so that $0<m<1$, and with wavelength given by
(\ref{eq2.12}). See Fig.2(a) for an example.
\begin{figure}[h]
\centering \subfloat[]{\label{fig:1}
\includegraphics[height=1.1in,width=1.2in]{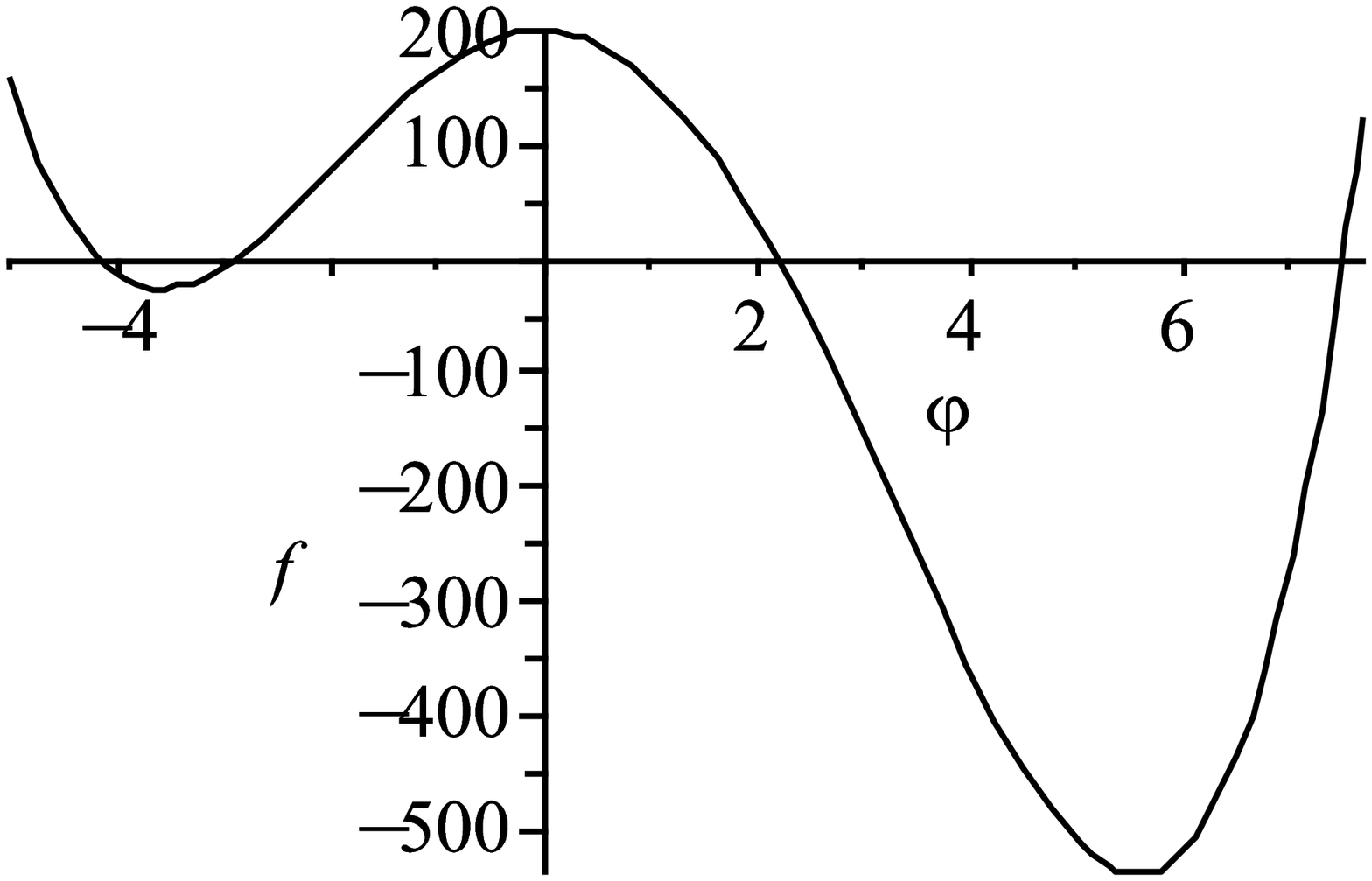}}\hspace{0.01\textwidth}
\subfloat[ ]{ \label{fig:2}
\includegraphics[height=1.1in,width=1.2in]{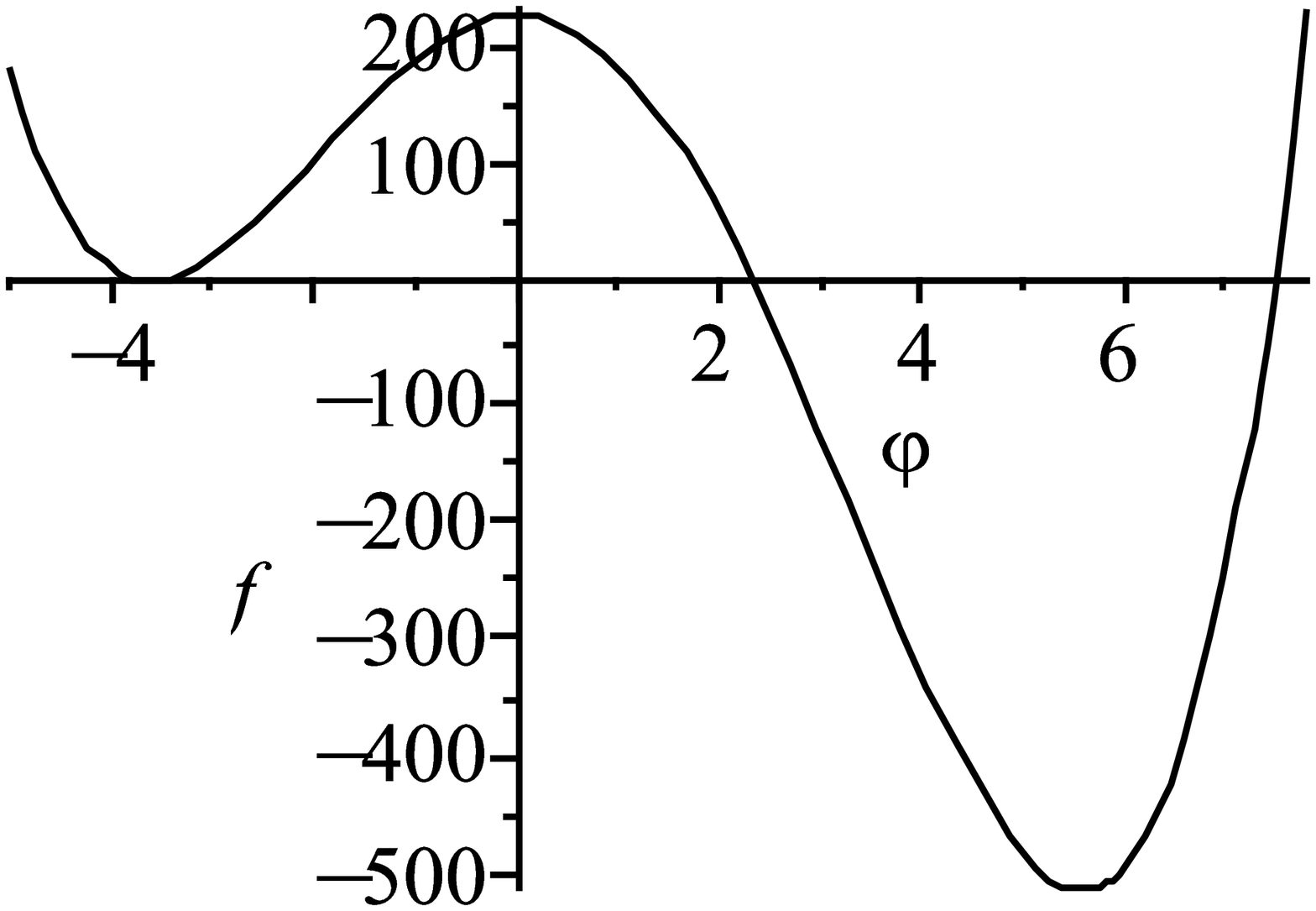}}\hspace{0.01\textwidth}
\subfloat[]{ \label{fig:3}
\includegraphics[height=1.1in,width=1.2in]{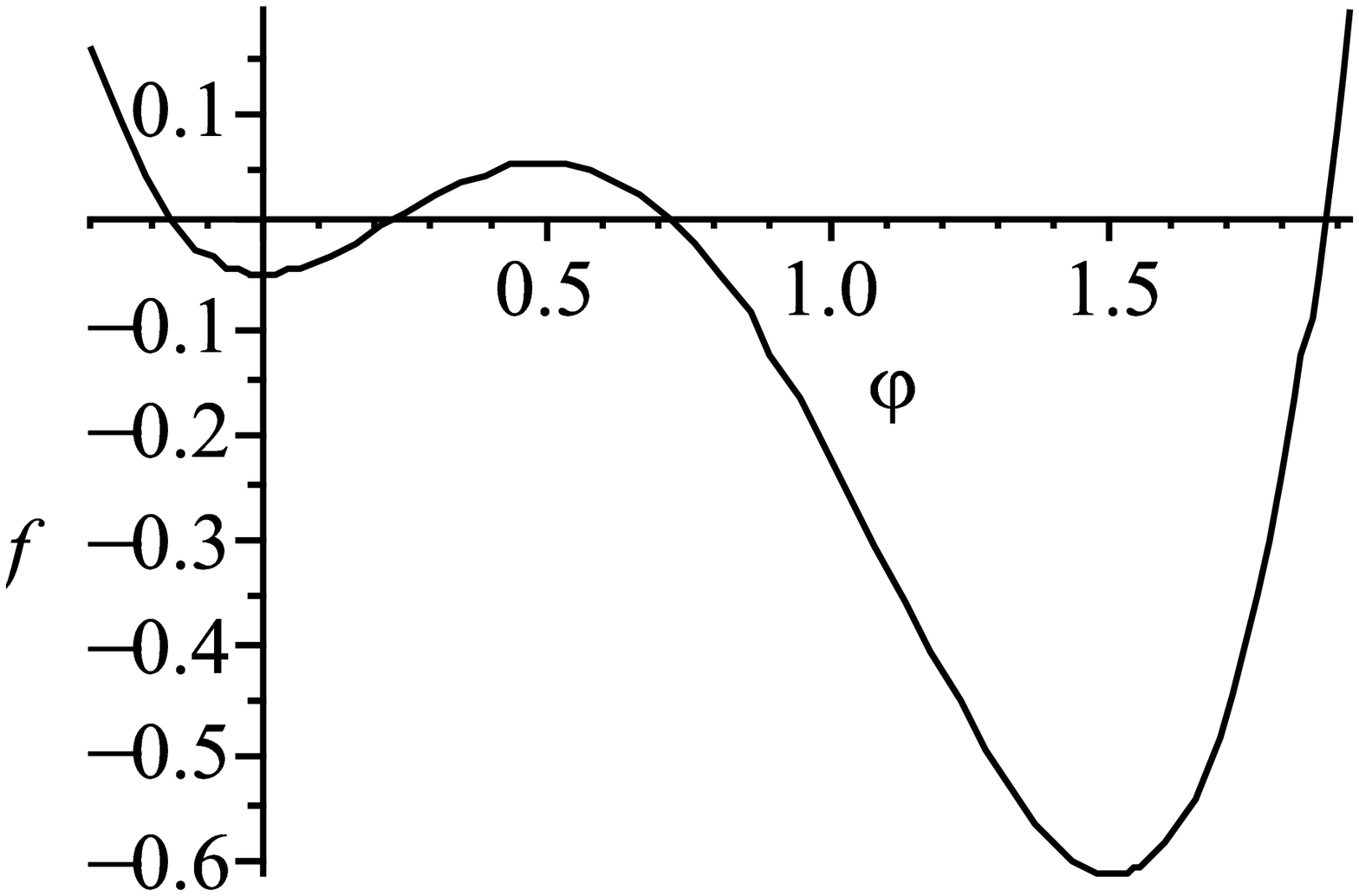}}\hspace{0.01\textwidth}
\subfloat[ ]{ \label{fig:4}
\includegraphics[height=1.1in,width=1.2in]{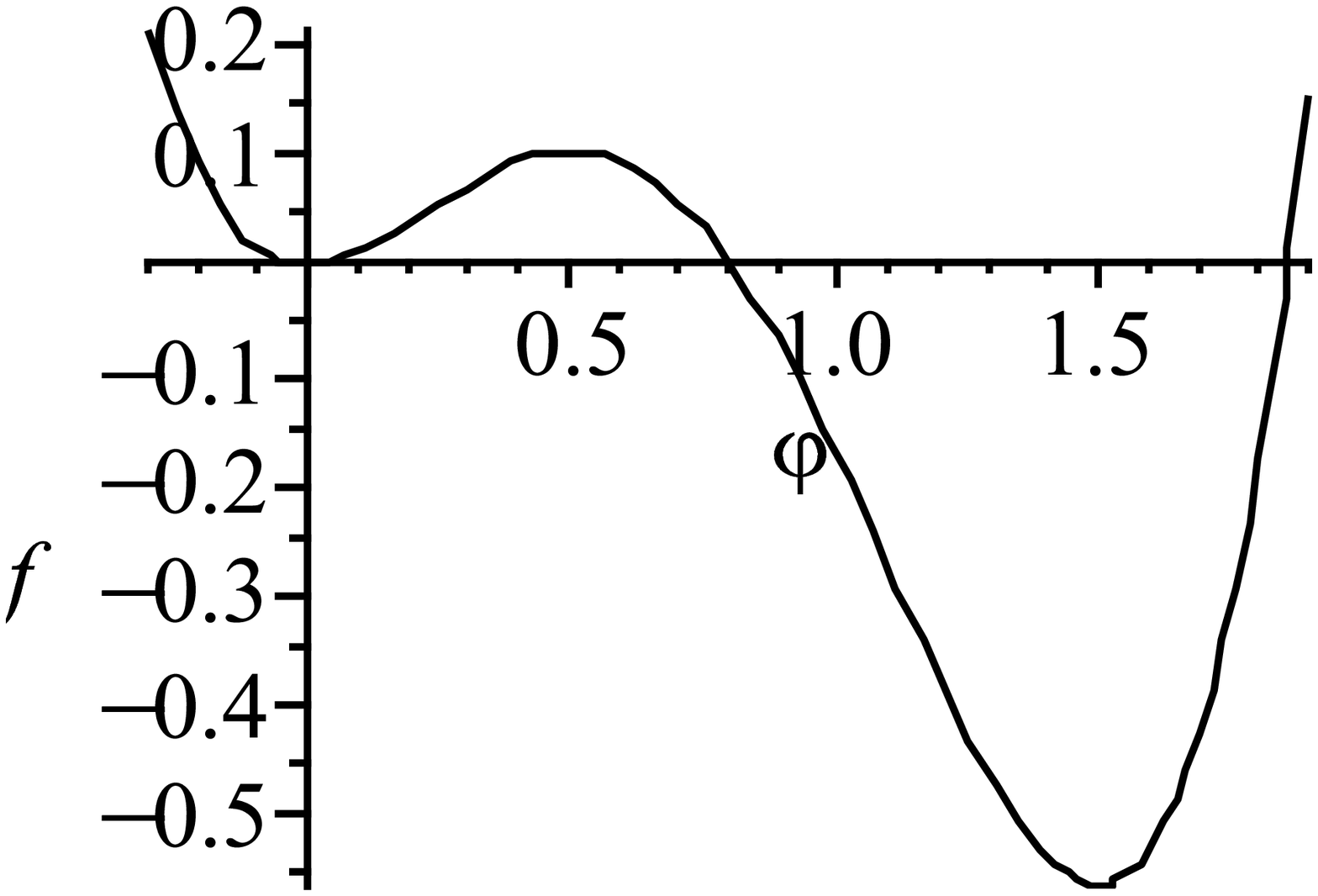}}\\
\subfloat[]{ \label{fig:5}
\includegraphics[height=1.in,width=1.2in]{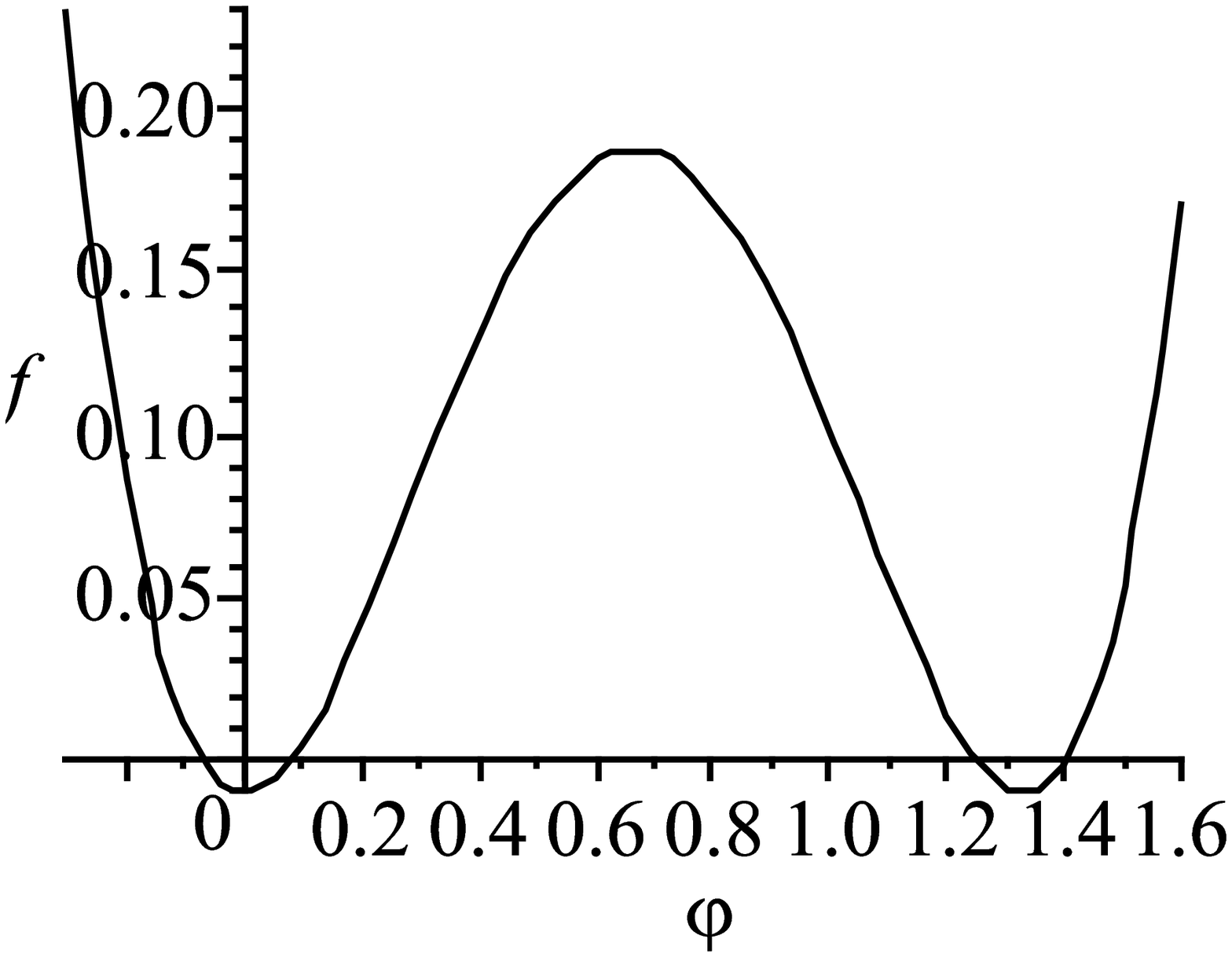}}\hspace{0.01\textwidth}
\subfloat[ ]{ \label{fig:6}
\includegraphics[height=1.in,width=1.2in]{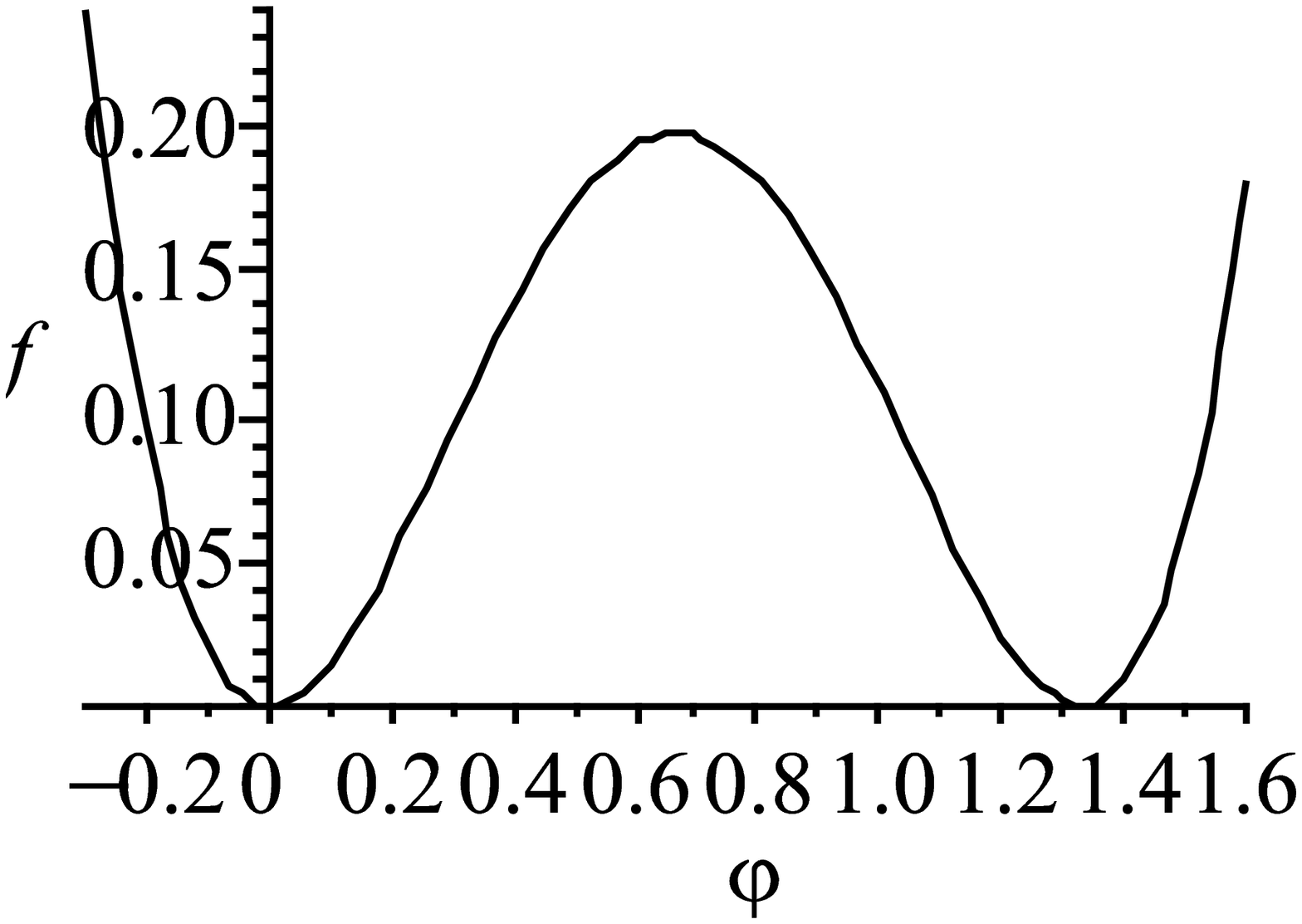}}\hspace{0.01\textwidth}
\subfloat[]{ \label{fig:7}
\includegraphics[height=1in,width=1.2in]{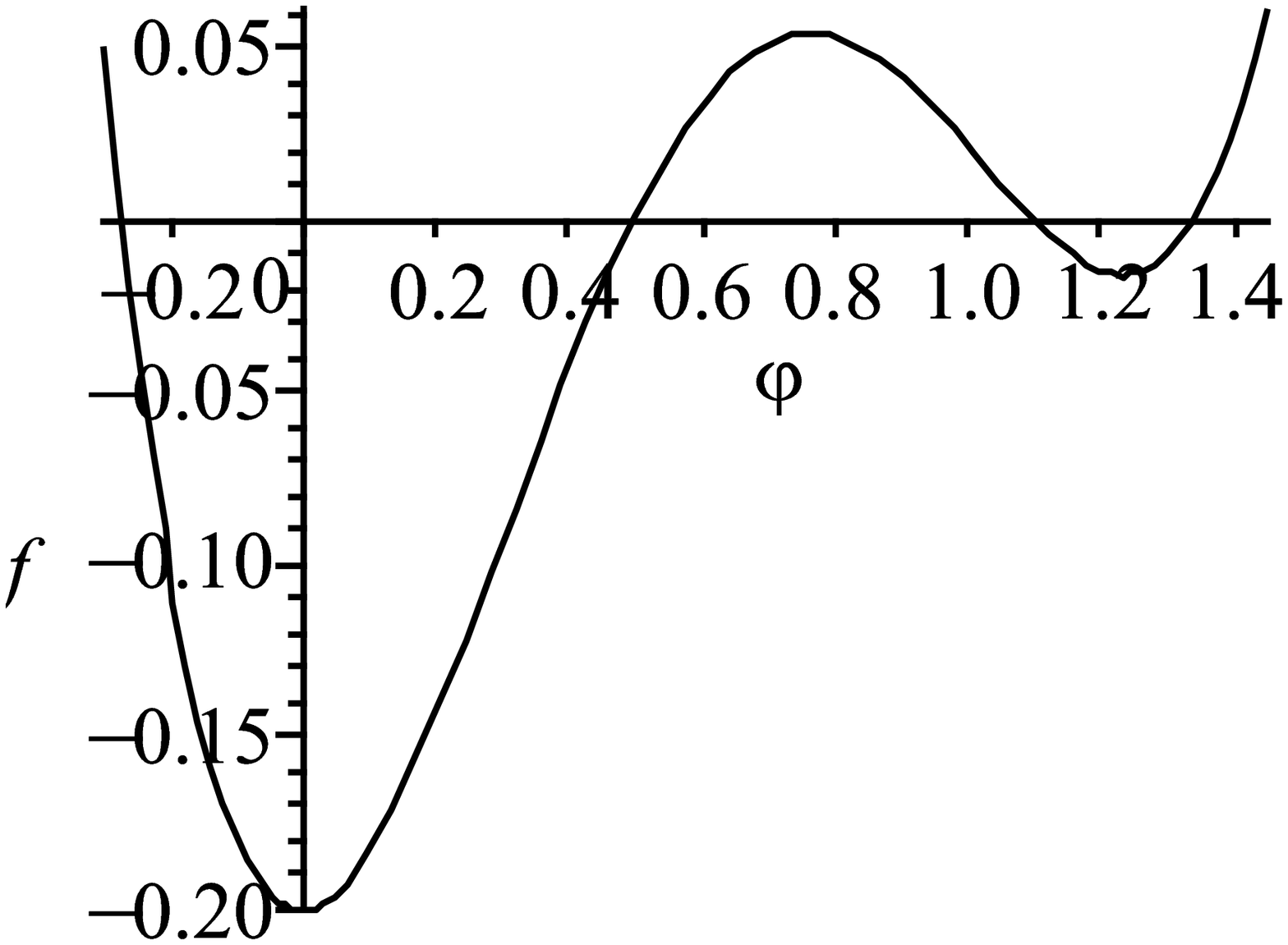}}\hspace{0.01\textwidth}
\subfloat[ ]{ \label{fig:8}
\includegraphics[height=1in,width=1.2in]{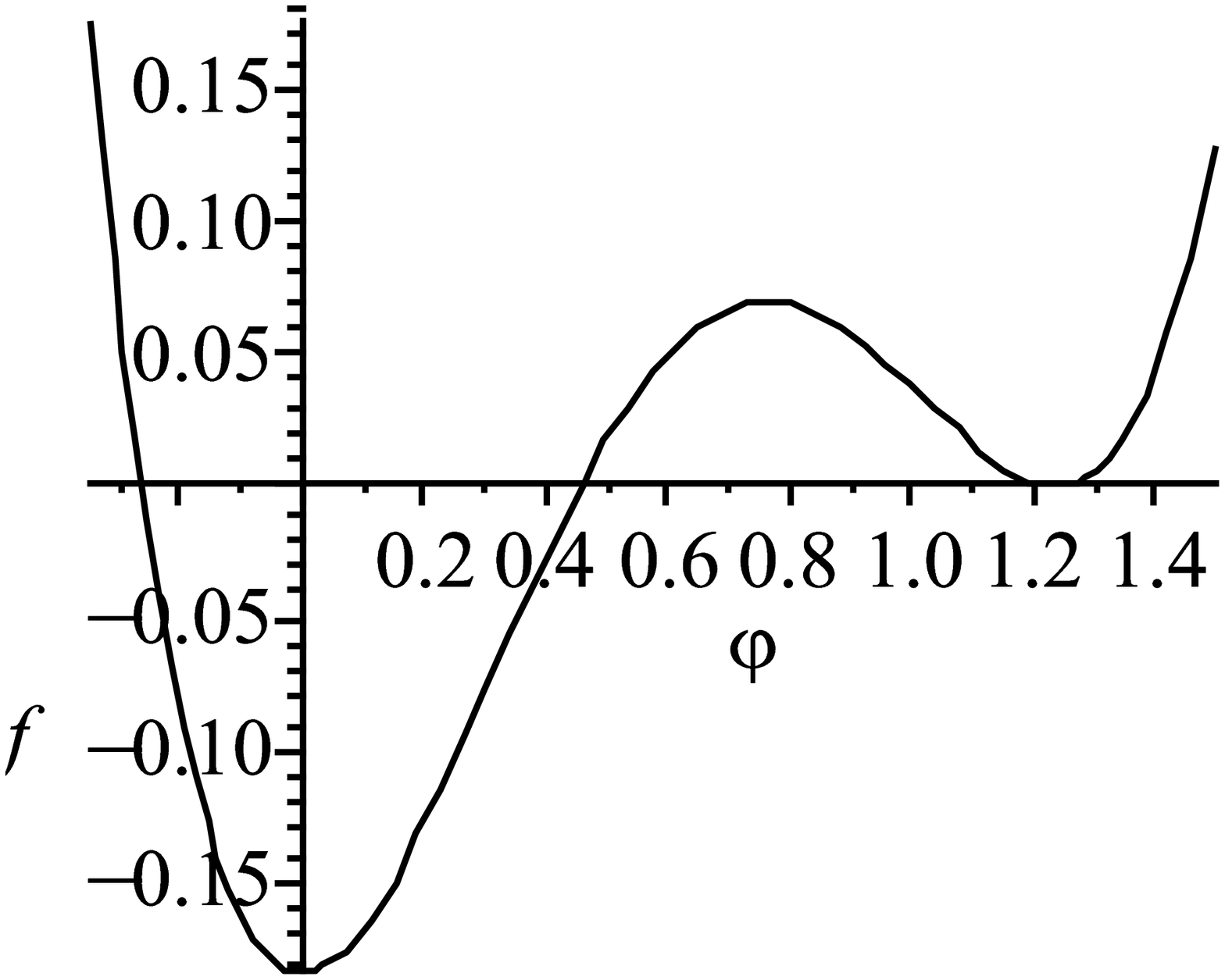}}

\caption{The curve of $f(\varphi)$ for the wave speed $c=2$. (a)
$a=-40$, $b=200$;
 (b) $a=-40$, $b=226.0424$;
 (c) $a=1.5$, $b=-0.05$;
 (d) $a=1.5$, $b=0$;
 (e) $a=\frac{\textstyle 16}{\textstyle 9}$, $b=-0.1$;
 (f) $a=\frac{\textstyle 16}{\textstyle 9}$, $b=0$;
 (g) $a=\frac{\textstyle 17}{\textstyle 9}$, $b=-0.24$;
 (h) $a=\frac{\textstyle 17}{\textstyle 9}$, $b=-0.1842$.}
\end{figure}

The case $a<0$ and $b=b_L$ (a corresponding curve of $f(\varphi)$ is
shown in Fig.1(b)) corresponds to the limit
$\varphi_1=\varphi_2=\varphi_L$ so that $m=1$, and then the solution
is a loop-like solitary wave given by (\ref{eq2.13}) with
$\varphi_2\leq \varphi <\varphi_R$ and
\begin{equation}
  \label{eq3.9}
  \varphi_3=\frac{\textstyle1}{\textstyle2}\sqrt{c^2-2a}+\frac{\textstyle c}{\textstyle6}-\frac{\textstyle1}{\textstyle3}
  \sqrt{c^2+3c\sqrt{4-2a}},
 \end{equation}
 \begin{equation}
  \label{eq3.10}
   \varphi_4=\frac{\textstyle1}{\textstyle2}\sqrt{c^2-2a}+\frac{\textstyle c}{\textstyle6}+\frac{\textstyle1}{\textstyle3}
  \sqrt{c^2+3c\sqrt{4-2a}}.
 \end{equation}
See Fig.3(a) for an example.

(2) $0<a<\frac{\textstyle 4 c^2}{\textstyle 9}$

In this case $0<\varphi_L<\varphi_R$ and $f(\varphi_R)<f(0)$. For
each value $0<a<\frac{\textstyle 4 c^2}{\textstyle 9}$ and $b_L<b<0$
(a corresponding curve of $f(\varphi)$ is shown in Fig.1(c)), there
are periodic valley-like solutions to Eq.(\ref{eq3.3}) given by
(\ref{eq2.10}) so that $0<m<1$, and with wavelength given by
(\ref{eq2.12}). See Fig.2(b) for an example.

The case $0<a<\frac{\textstyle 4 c^2}{\textstyle 9}$ and $b=0$ (a
corresponding curve of $f(\varphi)$ is shown in Fig.1(d))
corresponds to the limit $\varphi_1=\varphi_2=0$ so that $m=1$, and
then the solution can be given by (\ref{eq2.13}) with $\varphi_3$
and $\varphi_4$ given by the roots of $g(\varphi)=0$, namely
\begin{equation}
 \label{eq3.11} \varphi_3=\frac{\textstyle
2 c}{\textstyle 3}-\sqrt{\frac{\textstyle 4 c^2}{\textstyle 9}-a},
\varphi_4=\frac{\textstyle 2 c}{\textstyle 3} +
\sqrt{\frac{\textstyle 4 c^2}{\textstyle 9}-a}.
 \end{equation}
In this case we obtain a weak solution, namely the periodic
downward-cusp wave
 \begin{equation}
  \label{eq3.12}
  \varphi=\varphi(\xi-2j\xi_m), (2j-1)\xi_m<\xi<(2j+1)\xi_m, \
  j=0,\pm1, \pm2, \cdots,
 \end{equation}
where
\begin{equation}
  \label{eq3.13}
  \varphi(\xi)=(\varphi_3-\varphi_4\tanh^2(\xi/4))\cosh^2(\xi/4),
 \end{equation}
and
\begin{equation}
  \label{eq3.14}
  \xi_m=4\tanh^{-1}\sqrt{\frac{\varphi_3}{\varphi_4}}.
 \end{equation}
See Fig.3(b) for an example.

 \begin{figure}[h]
\centering \subfloat[]{\label{fig:1}
\includegraphics[height=1.in,width=2.4in]{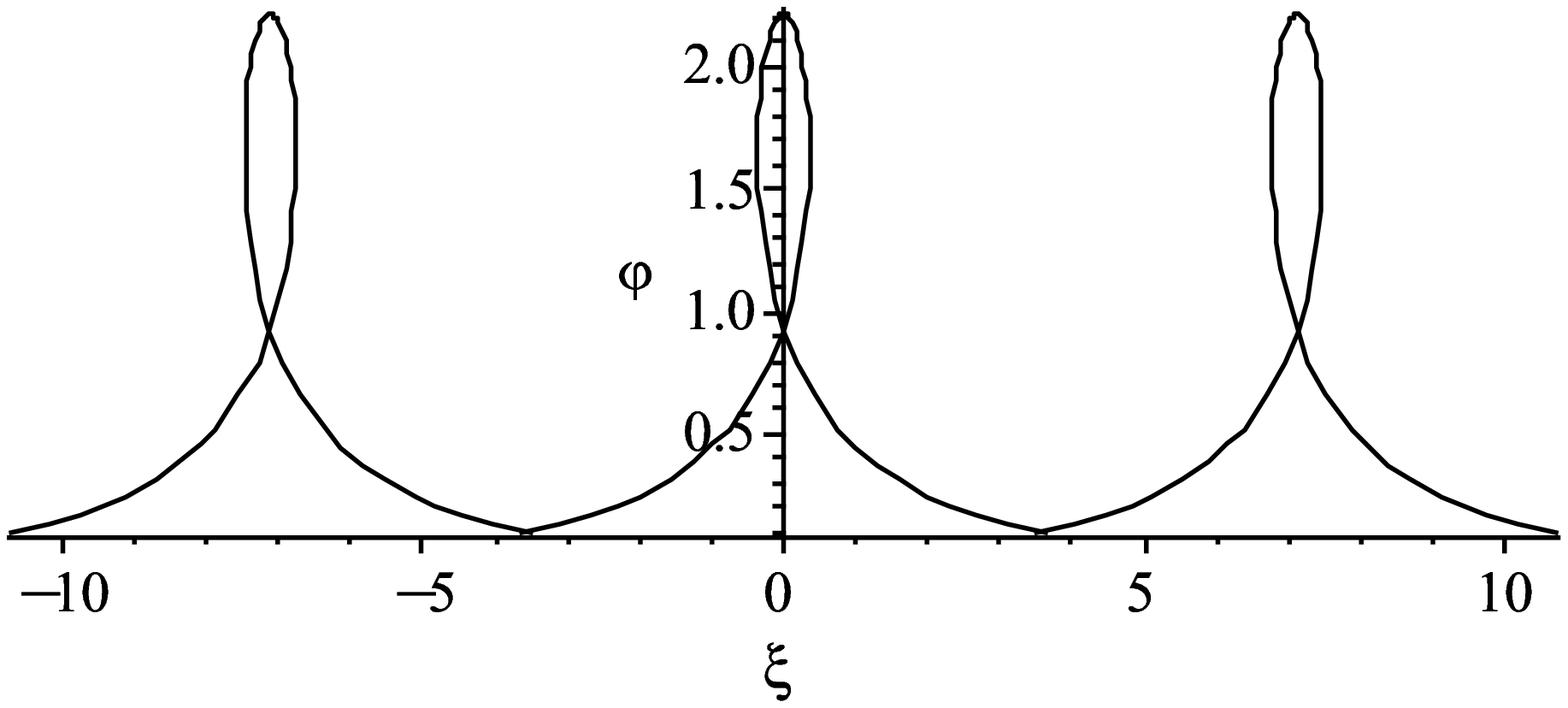}}\hspace{0.05\textwidth}
\subfloat[ ]{ \label{fig:2}
\includegraphics[height=1.in,width=2.4in]{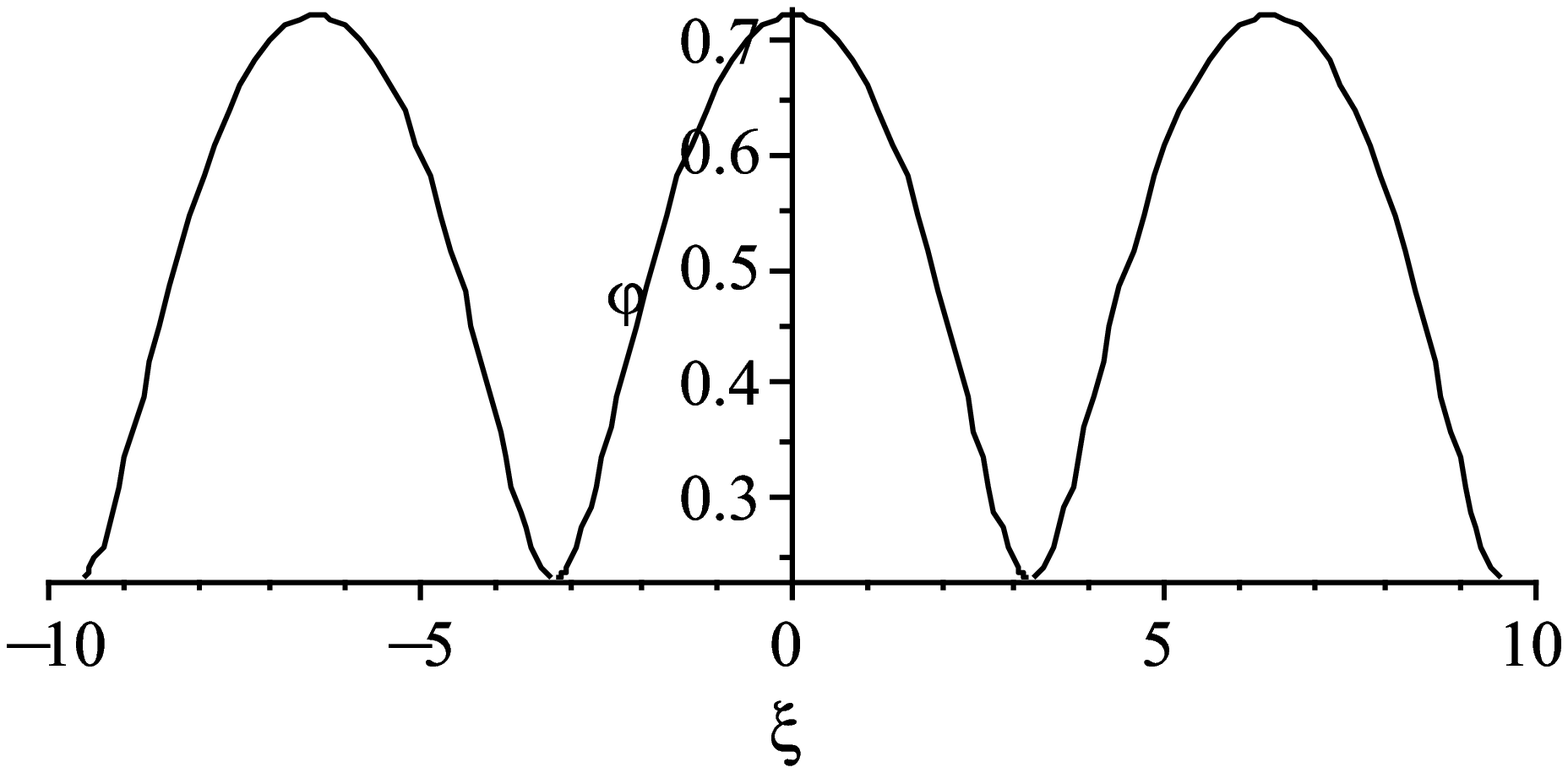}}\\
\subfloat[]{ \label{fig:3}
\includegraphics[height=1.in,width=2.4in]{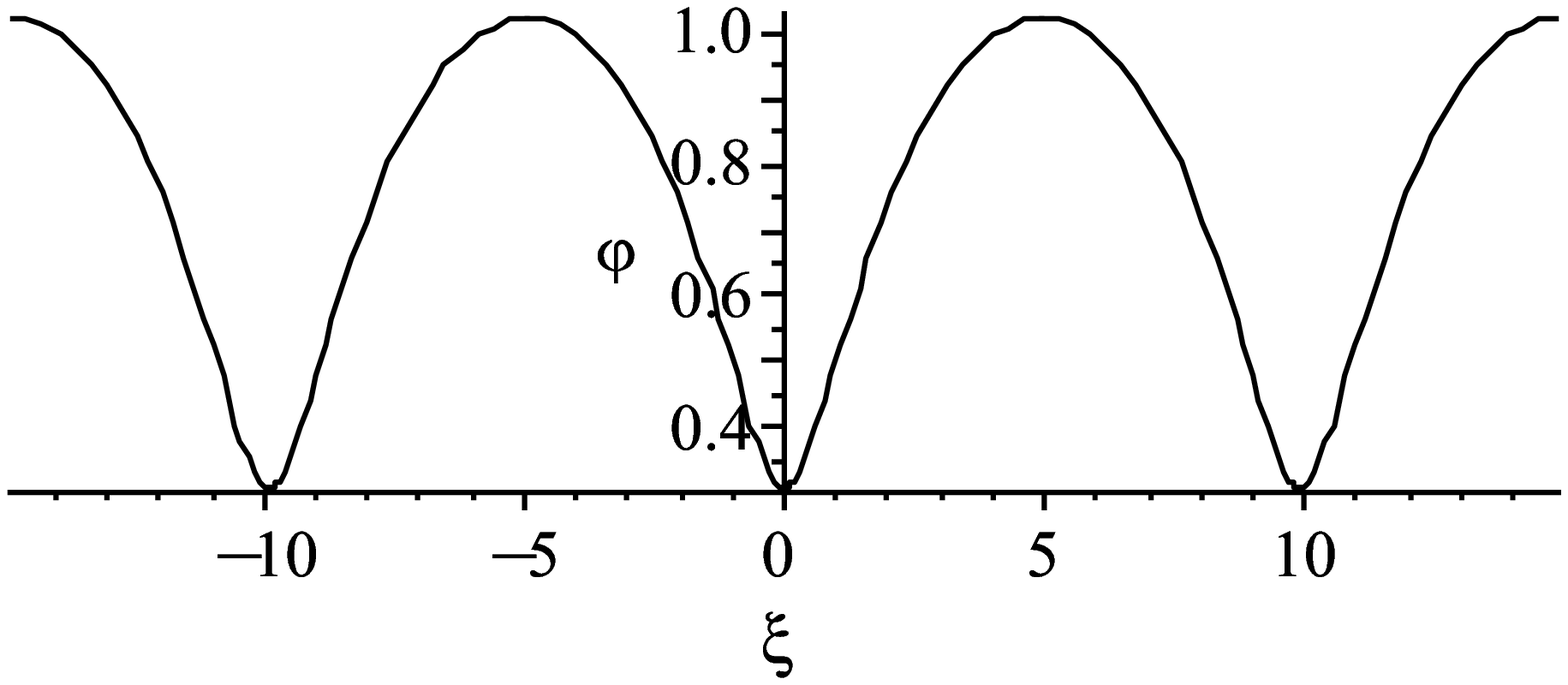}}\hspace{0.05\textwidth}
\subfloat[ ]{ \label{fig:4}
\includegraphics[height=1.in,width=2.4in]{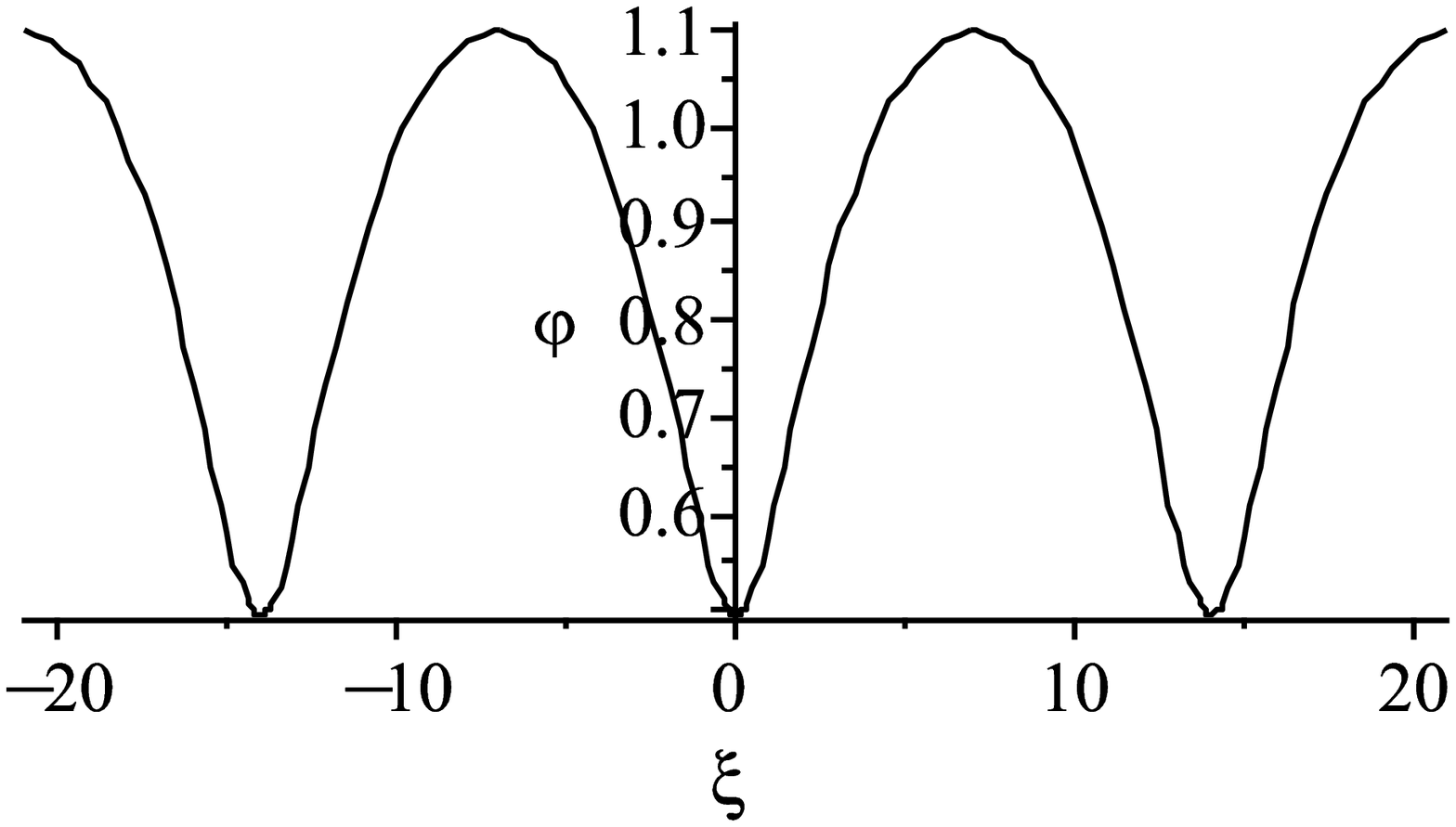}}

\caption{Periodic solutions to Eq.(\ref{eq3.3}) with $0<m<1$ and the
wave speed $c=2$.
 (a) $a=-40$, $b=200$ so $m=0.8978$;
 (b) $a=1.5$, $b=-0.05$ so $m=0.6893$;
 (c) $a=\frac{\textstyle 16}{\textstyle 9}$, $b=-0.1$ so $m=0.8254$;
 (d) $a=\frac{\textstyle 17}{\textstyle 9}$, $b=-0.24$ so $m=0.8412$.}
\end{figure}

 \begin{figure}[h]
\centering \subfloat[]{\label{fig:1}
\includegraphics[height=1.2in,width=2.2in]{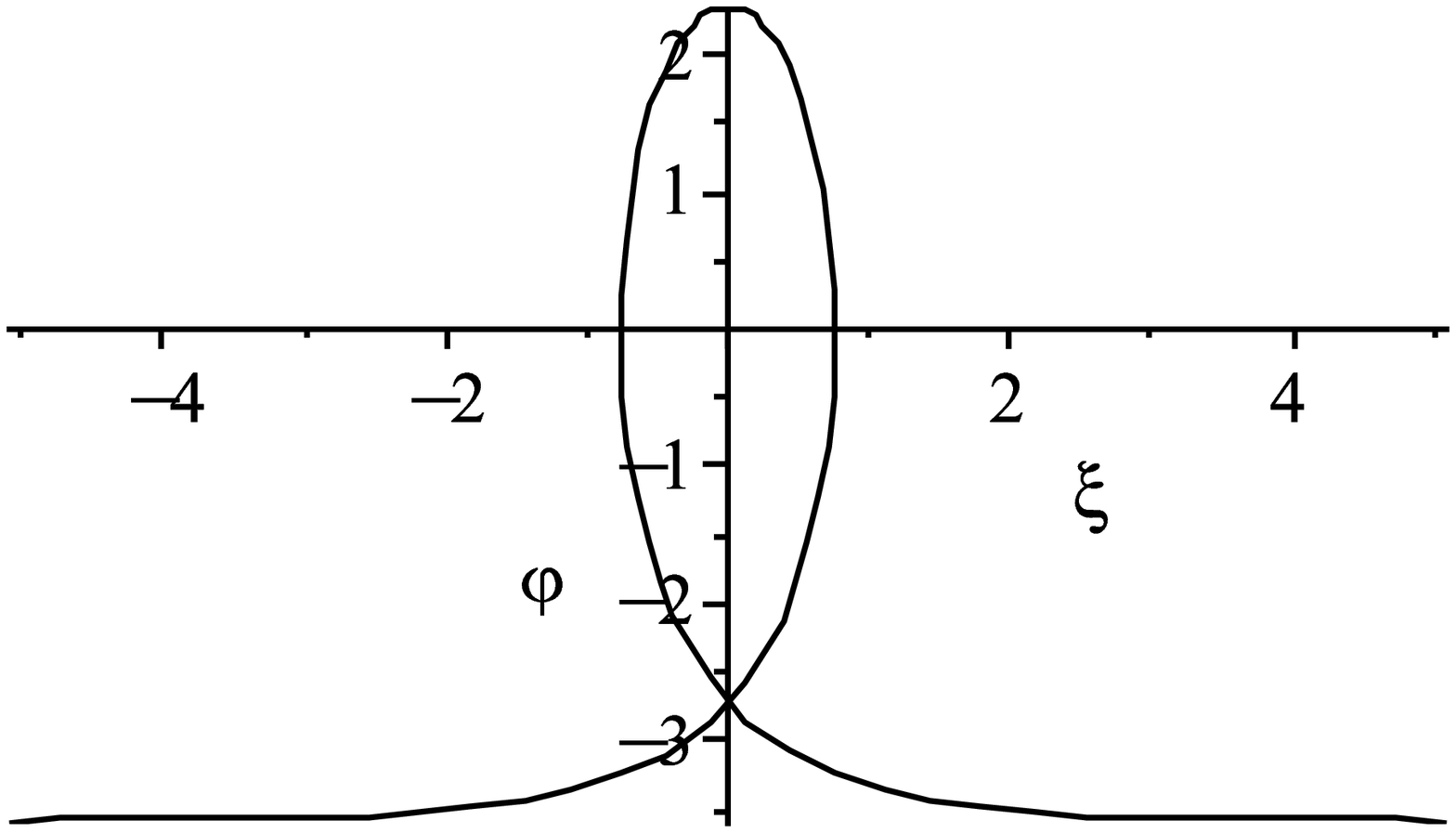}}\hspace{0.1\textwidth}
\subfloat[ ]{ \label{fig:2}
\includegraphics[height=1.2in,width=2.2in]{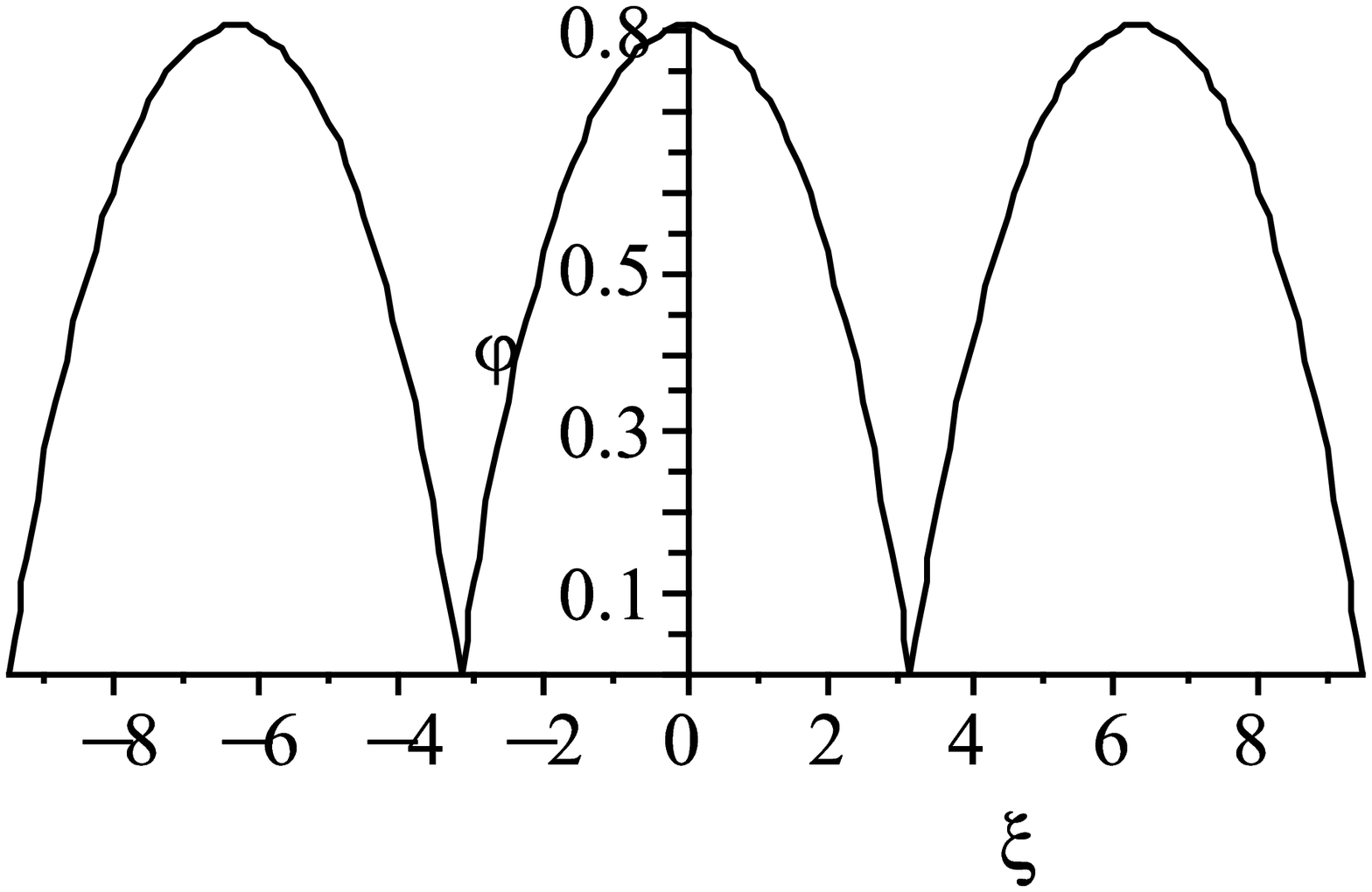}}\\
\subfloat[]{ \label{fig:3}
\includegraphics[height=1.2in,width=2.2in]{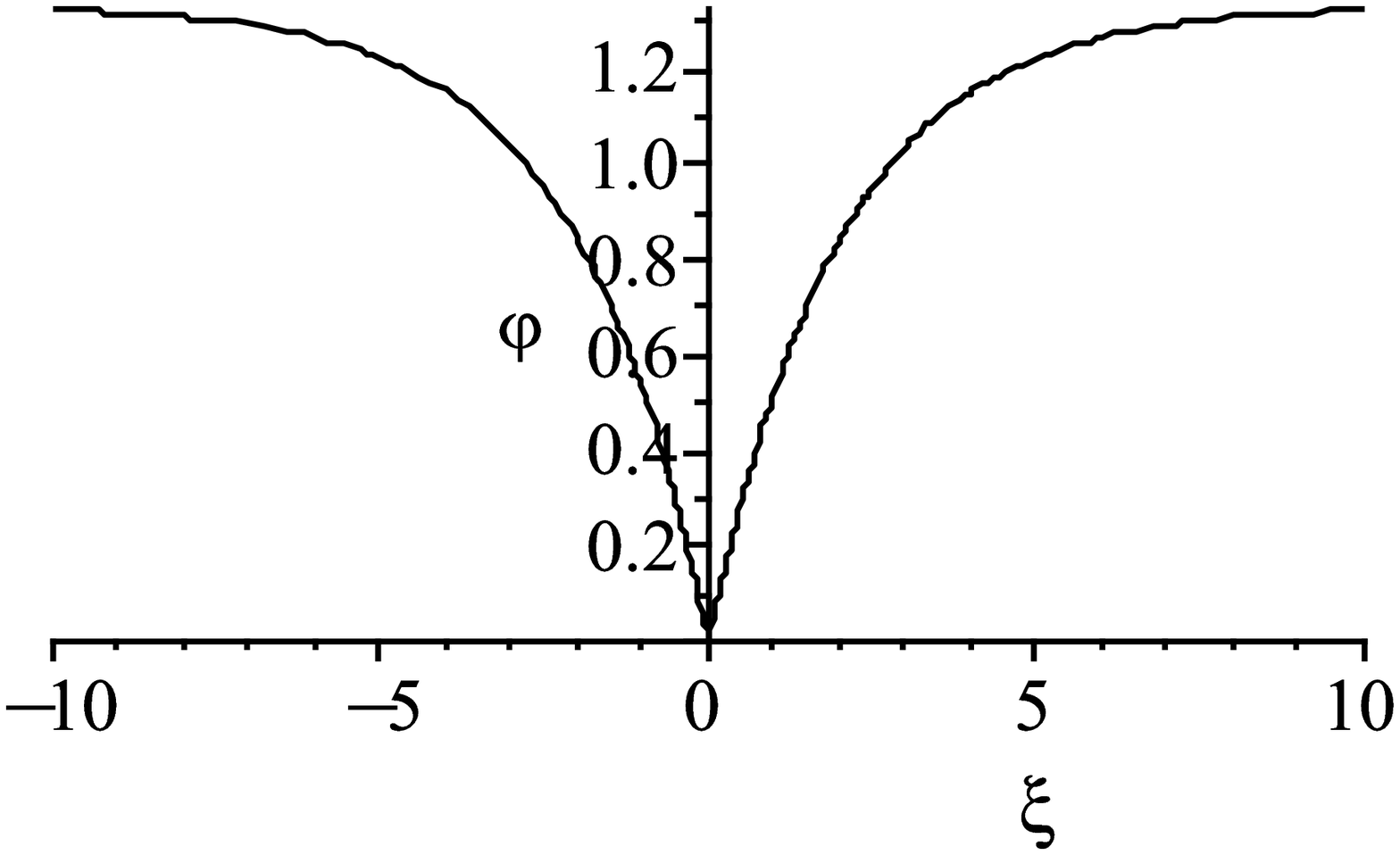}}\hspace{0.1\textwidth}
\subfloat[ ]{ \label{fig:4}
\includegraphics[height=1.2in,width=2.4in]{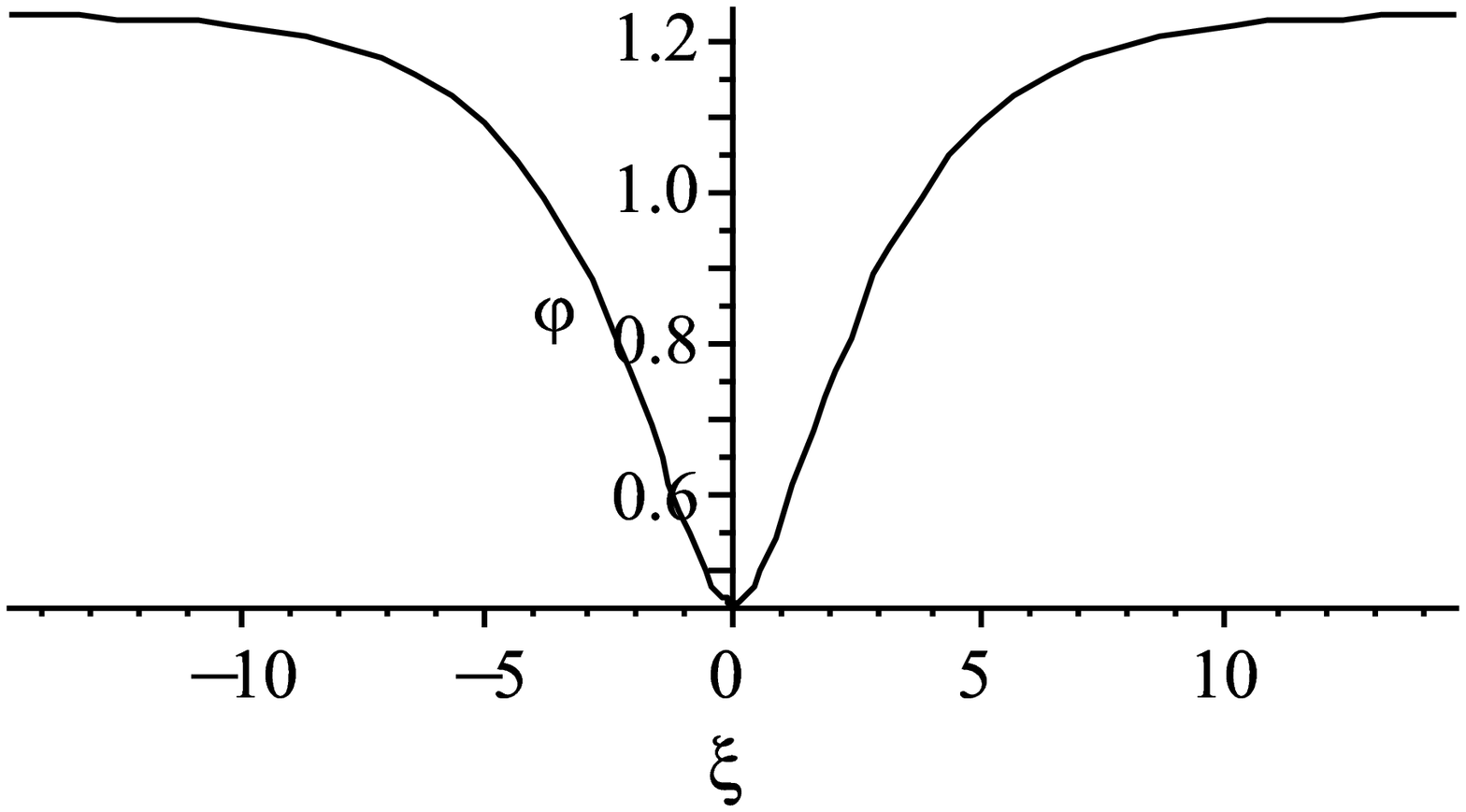}}
\caption{Solutions to Eq.(\ref{eq3.3}) with $m=1$ and the wave speed
$c=2$.
 (a) $a=-40$, $b=226.0424$;
 (b) $a=1.5$, $b=0$;
 (c) $a=\frac{\textstyle 16}{\textstyle 9}$, $b=0$;
 (d) $a=\frac{\textstyle 17}{\textstyle 9}$, $b=-0.1842$.}
\end{figure}

 (3) $a=\frac{\textstyle 4 c^2}{\textstyle 9}$

In this case $0<\varphi_L<\varphi_R$ and $f(\varphi_R)=f(0)$.  For
$a=\frac{\textstyle 4 c^2}{\textstyle 9}$ and each value $b_L<b<0$
(a corresponding curve of $f(\varphi)$ is shown in Fig.1(e)), there
are periodic valley-like solutions to Eq.(\ref{eq3.3}) given by
(\ref{eq2.5}) so that $0 < m < 1$, and with wavelength given by
(\ref{eq2.8}). See Fig.2(c) for an example.

The case $a=\frac{\textstyle 4 c^2}{\textstyle 9}$ and $b=0$ (a
corresponding curve of $f(\varphi)$ is shown in Fig.1(f))
corresponds to the limit
$\varphi_3=\varphi_4=\varphi_R=\frac{\textstyle 2 c}{\textstyle 3}$
and $\varphi_1=\varphi_2=0$ so that $m=1$. In this case neither
(\ref{eq2.9}) nor (\ref{eq2.13}) is appropriate. Instead we consider
Eq.(\ref{eq3.3}) with $f(\varphi)=\frac{\textstyle 1}{\textstyle
4}\varphi^2(\varphi-\frac{\textstyle 2 c}{\textstyle 3})^2$ and note
that the bound solution has $0<\varphi<\frac{\textstyle 2
c}{\textstyle 3}$. On integrating Eq.(\ref{eq3.3}) and setting
$\varphi=0$ at $\xi=0$ we obtain a weak solution
\begin{equation}
  \label{eq3.15}
  \varphi=-\frac{\textstyle 2 c}{\textstyle 3}\exp{(-\frac{\textstyle 1}{\textstyle 2}|\xi|)}+\frac{\textstyle 2 c}{\textstyle 3},
 \end{equation}
i.e. a single valley-like peaked solution with amplitude
$\frac{\textstyle 2 c}{\textstyle 3}$. See Fig.3(c) for an example.

 (4) $\frac{\textstyle 4 c^2}{\textstyle 9}<a<\frac{\textstyle c^2}{\textstyle 2}$

In this case $0<\varphi_L<\varphi_R$ and $f(\varphi_R)>f(0)$.  For
each value $\frac{\textstyle 4 c^2}{\textstyle 9}<a<\frac{\textstyle
c^2}{\textstyle 2}$ and $b_L<b<b_R$ (a corresponding curve of
$f(\varphi)$ is shown in Fig.1(g)), there are periodic valley-like
solutions to Eq.(\ref{eq3.3}) given by (\ref{eq2.5}) so that $0 < m
< 1$, and with wavelength given by (\ref{eq2.8}). See Fig.2(d) for
an example.

The case $\frac{\textstyle 4 c^2}{\textstyle 9}<a<\frac{\textstyle
c^2}{\textstyle 2}$ and $b=b_R$ (a corresponding curve of
$f(\varphi)$ is shown in Fig.1(h)) corresponds to the limit
$\varphi_3=\varphi_4=\varphi_R$ so that $m=1$, and then the solution
is a velley-like solitary wave given by (\ref{eq2.10}) with
$\varphi_L<\varphi\leq\varphi_3$ and
\begin{equation}
  \label{eq3.16}
  \varphi_1=\frac{\textstyle c}{\textstyle6}-\frac{\textstyle 1}{\textstyle 2}\sqrt{c^2-2a}
  -\frac{\textstyle 1}{\textstyle 3}\sqrt{c^2-3c \sqrt{c^2-2a}},
 \end{equation}
 \begin{equation}
  \label{eq3.17}
   \varphi_2=\frac{\textstyle c}{\textstyle6}-\frac{\textstyle 1}{\textstyle 2}\sqrt{c^2-2a}
 +\frac{\textstyle 1}{\textstyle 3}\sqrt{c^2-3c \sqrt{c^2-2a}}.
 \end{equation}
See Fig.3(d) for an example.

\section{Conclusion}

In this paper, we have found expressions for two types of traveling
wave solutions to the osmosis K(2, 2) equation, that is, the soliton
and periodic wave solutions. These solutions depend, in effect, on
two parameters $a$ and $m$. For $m=1$, there are loop-like ($a<0$),
peakon ($a=\frac{\textstyle 4c^2}{\textstyle 9}$) and smooth
($\frac{\textstyle 4c^2}{\textstyle 9}<a<\frac{\textstyle
c^2}{\textstyle 2}$) soliton solutions. For $m=1,
0<a<\frac{\textstyle 4 c^2}{\textstyle 9}$ or $0<m<1,
a<\frac{\textstyle c^2}{\textstyle 2}$ and $a \neq 0$, there are
periodic wave solutions.

\end{document}